# Beating the classical limit: A diffraction-limited spectrograph for an arbitrary input beam

Christopher H. Betters,[1,2*] Sergio G. Leon-Saval,[1] J. Gordon Robertson,[1,2] and Joss Bland-Hawthorn[1,2]

[1] *Institute of Photonics and Optical Science, School of Physics, University of Sydney, 2006, Australia*
[2] *Sydney Institute for Astronomy, School of Physics, University of Sydney, 2006, Australia*
[*]*c.betters@physics.usyd.edu.au*

**Abstract:** We demonstrate a new approach to classical fiber-fed spectroscopy. Our method is to use a photonic lantern that converts an arbitrary (e.g. incoherent) input beam into $N$ diffraction-limited outputs. For the highest throughput, the number of outputs must be matched to the total number of unpolarized spatial modes on input. This approach has many advantages: (i) after the lantern, the instrument is constructed from 'commercial off the shelf' components; (ii) the instrument is the minimum size and mass configuration at a fixed resolving power and spectral order (~shoebox sized in this case); (iii) the throughput is better than 60% (slit to detector, including detector QE of ~80%); (iv) the scattered light at the detector can be less than 0.1% (total power). Our first implementation operates over 1545-1555 nm (limited by the detector, a 640×512 array with 20μm pitch) with a spectral resolution of 0.055nm ($R$~30,000) using a 1×7 (1 multi-mode input to 7 single-mode outputs) photonic lantern. This approach is a first step towards a fully integrated, multimode photonic microspectrograph.

## 1. Introduction

One of the fundamental design characteristics of any spectrograph is the amount of light collected, dispersed and re-imaged onto a detector. This quantity is characterized by the étendue (or AΩ factor, defined as the product of entrance/collection surface area, A, and the solid angle of acceptance of the optical system, Ω). Conventionally, in order to maximize the throughput (ratio of photons detected to photons collected) of a spectrograph, the étendue of all optical elements (e.g. from a telescope to the detector) must be well matched.

The most challenging spectrographs are those that operate at high spectral resolution ($R = \lambda/\Delta\lambda > 20,000$). When observing faint sources, there is a strong tension between the need to broaden the slit to allow for more light from the source while preserving the spectroscopic performance of the instrument. An additional challenge is that for a spectrograph operating at a resolving power $R$ matched to natural seeing (i.e. a fixed angular resolution limit measured in arcseconds), the size of the instrument scales in proportion to the telescope diameter [1]. Modern high-resolution astronomical spectrographs are already meter-scale instruments and are required to be even larger on the next generation of extremely large telescopes (25-40 m diameter). This is the impasse reached in the design of high-resolution spectrographs; maximizing collection area requires the physical size to increase in order to maintain a given resolution. Adaptive optics (AO) is one route to mitigate this problem, reducing the extent of an image on the spectrograph slit. In the visible region, the blue end in particular, current AO techniques are somewhat limited. However, in any wavelength range, the spectrograph designs are still paired with a given telescope/AO system.

Here we raise two key challenges: (i) to get more light through a high-resolution spectrograph; and (ii) to break the dependence of spectrograph size on the telescope diameter or other input. We solve both of these problems simultaneously through our new approach to spectrograph design. The basic principle is to remap the extent of the entrance slit, minimizing the width while conserving the effective étendue. The result is a long thin slit (ideally diffraction-limited in width) replacing the comparatively short wide classical slit. In principle, this is exactly what an integral field unit (IFU) does when reducing a 2D image to a linear slit. However, the devices used in conventional IFUs can already be extremely complex, so requiring they remap to even smaller sizes will invariably lead to losses of efficiency or increased cost (if not both). A new approach is required to efficiently reduce the slit to a diffraction-limited width.

In [1], Bland-Hawthorn et al. introduced the concept of a photonic integrated multimode microspectrograph (PIMMS) that promised a fundamental shift in spectrograph design which we now demonstrate for the first time. The proposal harnessed an emerging technology, known as the photonic lantern (PL), to allow efficient conversion of an essentially arbitrary input (e.g. a combination of arbitrarily excited modes) to the consistent and diffraction-limited format of single-mode fibers (SMFs) [2-5]. The PL remaps the entrance slit of the spectrograph, thus achieving throughput equivalent to a multimode fiber (MMF) design and the spectral resolution of a diffraction-limited slit width.

The PL's conversion from MM to SM completely decouples the spectrograph design from the light source at the MM input, resulting in a spectrograph design that no longer need match the input beam of a specific telescope (or any other source for that matter). Instead, it is designed to match the output of the array of SMFs (whose output remains fundamentally unchanged regardless of the source at the MM input). The conversion also allows single-mode photonic technologies to be incorporated into new and pre-existing spectrographs, the most successful so far being fiber Bragg gratings for sky OH suppression [6, 7].

In this paper, we will briefly describe the design considerations and figures of merit for a SMF fed spectrograph, including an overview of the 1×7 (1 MM to 7 SM) PL used. This is followed by a laboratory characterization of the prototype spectrograph, dubbed PIMMS IR.

## 2. Diffraction-limited design considerations

*2.1 Entrance slit*

The beam emerging from an SMF is diffraction-limited by definition [1] (i.e. the étendue or $A\Omega = \lambda^2$), with the light propagating in the LP01 mode (Gaussian like). Coupling to a SMF efficiently is not trivial. When coupling in ideal circumstances (i.e. a uniformly illuminated circular pupil), the efficiency peaks at 80% and decreases proportionally to the Strehl ratio. A typical example is coupling light from a telescope where, in the presence of atmospheric turbulence, the focus can be extremely perturbed, and the Strehl ratios are typically less than <1% and thus couple poorly to a sole SMF [8, 9]. The PL alleviates the coupling concern while simultaneously allowing the spectrograph's entrance slit to be formed by an array of SMF.

A factor to consider when forming the entrance slit in this way is the distance between the SM cores. If the fibers are placed in a simple linear array to form a "pseudo-slit," there will be a significant gap between spectra at the detector. This gap between cores (typically 125μm for standard SMF) leads to wasted detector space unless some additional component is used to bring the cores closer together. Ultrafast laser inscription of waveguides offers a path to do just that, with appropriate devices already demonstrated [10]. However, the addition of components can lead to further complication and become an additional source of aberration in a design intended to be as simple as possible. Another solution is to form the pseudo-slit with a hexagonal bundle of SMFs or a multi-core fiber, constructed and positioned such that each core forms an independent spectrum when dispersed [11]. However, for simplicity (and as it does not significantly affect the actual spectrograph design), we will assume in later sections that the SMF cores are aligned to form a linear pseudo-slit with an optimal spacing unless otherwise specified.

*2.2 Throughput/Étendue of a photonic lantern*

The fundamental requirement of the PL is that it efficiently transforms MM light into SM waveguide outputs. Currently the most common method of fabricating a PL starts with a bundle of SMFs inserted in a low-index glass capillary. Those are fused and drawn down together to form a composite tapered waveguide that ends in a MM port. At the MM end, the SMFs cladding material becomes the core and the low index glass capillary becomes the cladding. An alternative is to start with a multi-core fiber (MCF; i.e. a 2D array of SM cores in a single fiber), which are again drawn down and tapered within a glass capillary. In both cases, the cores of the SMFs become too small to guide light [2-5]. Another promising method is directly writing waveguides using ultrafast laser inscription, forming a PL in a solid block of glass[12, 13].

In either method, if the taper transition from MM to SM is optically adiabatic, the modes of the MMF input will evolve into the modes of the SM outputs. In order for the conversion to be efficient, the number of modes supported in the MM ($m$) port should match the number of SM fibers ($n$). The number of modes in a step-index MMF is determined from the core size and the acceptance angle (specified as a numerical aperture, NA) or in terms of the fiber V parameter[14],

$$m \approx \frac{V^2}{4} \approx \left(\frac{\pi d}{2\lambda}\text{NA}\right)^2, \qquad (1)$$

where $d$ is the fiber core diameter and $\lambda$ is the free space wavelength of the guided light. It should be noted that each mode is capable of carrying two polarizations. Consequently when polarization is considered there are effectively twice as many modes. Even so, the number of SMFs remains the same as each fiber can also carry two polarizations.

For low loss operation, the number of modes has to be conserved (i.e. $m=n$) at the PL's design wavelength. A typical throughput of the current generation of lanterns is ~93% (0.3dB loss) [5]. Note that as $\lambda$ decreases the number of modes in the MM port increases. This will result in an effective reduction in throughput blueward of the design wavelength when the number of modes in the MM port exceeds the number of modes available in SM outputs. In contrast, redward of the design wavelength, the number of SM outputs will exceed the number of modes supported in the MM port, thus the transition will not have a negative impact on throughput. So, an absolute limit on PL transmission throughput can simply be expressed by the ratio of the number of SM outputs and number of modes in the MM input, i.e.

$$\text{transmission fraction} = \frac{n}{m} = \frac{n(A\Omega)_{\text{SMF}}}{(A\Omega)_{\text{MMF}}}. \qquad (2)$$

The number of modes can be thought of as equivalent to the étendue of fiber, indeed, simply multiplying by $\lambda^2$ returns the étendue for low numerical apertures.

*2.3 Resolving power*

Another key figure of merit for a spectrograph is its minimum resolvable wavelength difference or resolution ($\Delta\lambda$), often expressed as the resolving power ($R = \lambda/\Delta\lambda$). This is a convolution of the instrumental point spread function (PSF; i.e. the grating resolution and imaging quality) and the width of the imaged entrance slit. In conventional spectrographs the slit width is generally the limiting factor [15]. Using the PL, the width of the slit can be brought to an absolute minimum leaving the instrumental PSF as the dominant factor in determining resolution. With diffraction-limited optics (without significant beam truncation) the instrument PSF should reproduce the near-field output of the SM fiber, and instrument resolution is then dependent only on the resolution of the diffraction grating.

For the case of a truncated Gaussian with a $1/e^2$ width, $d$, and angle of incidence, $\theta$, the resolving power is thus

$$R = \frac{\lambda}{\Delta\lambda} = 2.38 \frac{Kd}{\lambda} \tan\theta. \qquad (3)$$

where a generalized form of the Rayleigh criterion, where the resolution element is 1.112 × FWHM, was used in Eq. (3) to provide a realistic estimate of the resolving power [16, 17]. The factor $K$ accounts for the broadening effect seen when bringing a collimated Gaussian beam that has been truncated to a focus [17-19].

*2.4 Detector area implications of the photonic lantern*

The detector area required for a PIMMS type spectrograph is effectively the same as that required by equivalent MM spectrograph. This unexpected result requires some discussion and is illustrated in Fig. 1. Assume that the PIMMS and MM spectrographs share equal bandwidth and resolving power. The PIMMS case has an increase in the spatial extent of the spectrum (i.e. pseudo-slit length) that proportional to $\sqrt{N}$. This is balanced by a $\sqrt{N}$ reduction of space required in the wavelength direction (i.e. the PIMMS resolution element is related to the width of the MM case by a factor $1/\sqrt{N}$). So when determining the total area, the factor $\sqrt{N}$ is eliminated, with two qualifications. Firstly, the actually area required by the PIMMS design is dependent on the spacing of the SM cores at the pseudo-slit, but the spacing can be designed such that there is no difference. Secondly, while the detector areas are equal, the PIMMS case requires smaller pixels to adequately sample the PSF of the SM pseudo-slit. Indeed, it is clear that the pixels must be at least a factor $1/\sqrt{N}$ times smaller then those required for the MM case (as we must have same number of resolution elements in the wavelength axis for both the PIMMS and MM case). This indicates that PIMMS IR needs $N$ more pixels then the MM case. This assumes that the spatial extent needs the same degree of sampling (i.e. pixels per FWHM) as the wavelength axis, which is not necessarily the case. One could conceive a detector for PIMMS IR that has rectangular pixels, where the long axis in aligned with the spatial extent of the slit, and the short edge matched to adequately sample a resolution element. In such a case the number of extra pixels required is $\varepsilon.N$, where $\varepsilon$ is the ratio of SM sampling and MM sampling.

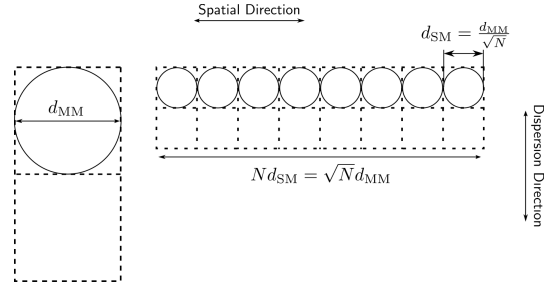

Fig. 1 – Illustrated are the images of a MM spectrograph slit (left) and the pseudo-slit of a PIMMS spectrograph (right) where both PSFs are sampled by 1 pixel. The relationship between $d_{SM}$ and $d_{MM}$ is found by equating the étendue. We can see a √N decrease in the wavelength direction width is balanced by a √N increase in width in the spatial direction, thus conserving area. In the case of equal sampling it is clear that we require N more pixels in the PIMMS case.

## 3. PIMMS IR – A single-mode fiber fed spectrograph

### 3.1 Design

The key components of the PIMMS design are illustrated in Fig. 2. Light is coupled into the multimode port of the 1×7 PL, where it is converted to an array of SMFs. The SMFs form the pseudo-slit for a compact and diffraction-limited spectrograph, here dubbed PIMMS IR. In the PIMMS IR design we use a collimator and camera combination with effective magnification of 7.7. The collimated beam has a $1/e^2$ diameter of 13mm and is dispersed by a volume phase holographic grating manufactured by Wasatch Photonics, Inc. with 1120 lines/mm. The primary detector is a Xenics Xeva series 320×256 InGaAs array with a pixel pitch of 30 microns (9.6mm × 7.68mm). We also use an alternate model with a 640×512 array with 20 microns pixels. The other optical components are drawn from the Thorlabs, Inc. catalogue.

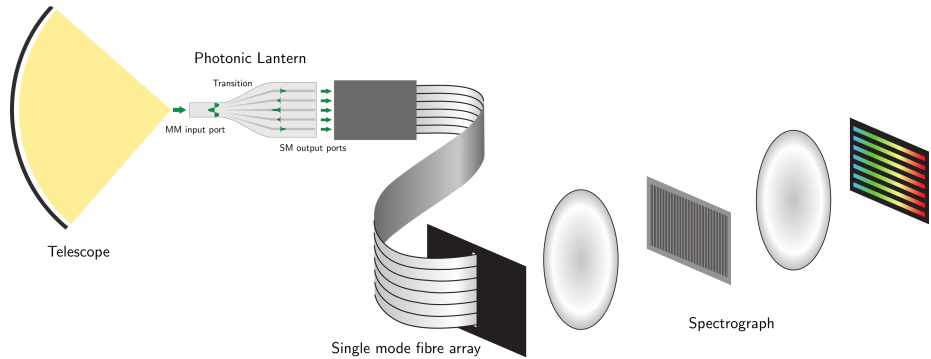

Fig. 2 - Schematic of PIMMS IR setup. Light form any source, in this case a telescope, is coupled to the MM port of a PL The PL produces multiple SMFs that form the pseudo-slit of a compact diffraction-limited bulk optic spectrograph.

### 3.2 Instrumental PSF

To characterize the instrumental PSF, we fed the PIMMS IR spectrograph with one SMF. The input was coupled directly from a tunable laser source (Tunics 3642 HE L) set to 1550.25 nm. A typical line profile (collapsed in the spatial direction with a Gaussian weighted summation) is shown in Fig. 3, along with the best-fit Gaussian, ZEMAX physical optics propagation (POP) simulations and the same laser line measured using an OSA (ADVANTEST Q8384). We also show the typical spectral line profiles attributed to the IRIS2 spectrograph [20] and the AAOmega spectrograph [21]. These show that the wings in a PIMMS design are much weaker than in a conventional design, besting IRIS2 by two orders of magnitude and AAOmega by one. Both IRIS2 and AAOmega have lower resolutions than PIMMS IR, and operate over different wavelength ranges. To provide a fair comparison they have been scaled to have the same FWHM, measured by their respective best-fit Gaussian, as PIMMS IR.

The measured FWHM of the PIMMS IR PSF of the spectral and spatial profiles is ~57.5μm (1.9 pixels). This is just 3.6% larger then that predicted by the POP simulation. In the spectral profile (top part of Fig. 3) the wings are remarkably consistent with those predicted by ZEMAX POP simulations. At low intensity levels, they do begin to depart from the simulation. The wings of the spatial profile (bottom part of Fig. 3) are slightly better than the spectral profile, but not as low as expected from the POP simulation. We note that in our simulations, if the truncation factor is increased (i.e. the diameter of lens is reduced) in the collimator the wings in the spatial profile will increase until they are approximately symmetrical with the spectral profile, at which point both the spectral and spatial wings increase in strength together. Thus, the symmetry of the measured PSF seems to indicate a broadening effect not fully accounted for in POP is the dominant cause of the discrepancies. We attribute this to the fact that (i) surface irregularities ($\lambda/4$ at 633 nm or $\sim\lambda/10$ at 1550 nm) are on the edge of diffraction-limited, which is not accounted for in POP; and (ii) slight differences in the alignment between ZEMAX and reality (e.g. a small change in the camera element separation results in an order of magnitude change in focal length). These factors are all inherent drawbacks of COTS components, which nonetheless perform extremely well.

The FWHM of 1.9 pixels corresponds to ~49 pm spectrally. So the resolving power using the generalized Rayleigh width (Eq. 3) is R~28,300. More than 99% of the total power is concentrated in the core of the PSF where it remains Gaussian-like. Less than 1% of the power is contained in the extended wings and is in part attributable to veiling glare from the bulk COTS components. This allows the design to resolve at higher resolutions in high contrast situations (assuming an appropriate detector of course). Table 1 summarizes the maximum dynamic range (i.e. intensity contrast between peaks) possible at spectral resolutions of 1, 2, 4 and 8 times the best resolution of PIMMS IR. For example two spectral lines that are separated by 0.11nm are resolved if the ratio between their peak intensities is not larger then 17.5dB (57:1).

Table 1. Maximum resolution attainable at various contrast levels with the PIMMS IR spectrograph.

| No. Resolution elements | Dynamic Range (dB) |
|---|---|
| 1 (0.0548 nm) | 0 (1:1) |
| 2 (0.110 nm) | 17.5 (57:1) |
| 4 (0.219 nm) | 42 (16500:1) |
| 8 (0.438 nm) | 51 (130,000:1) |

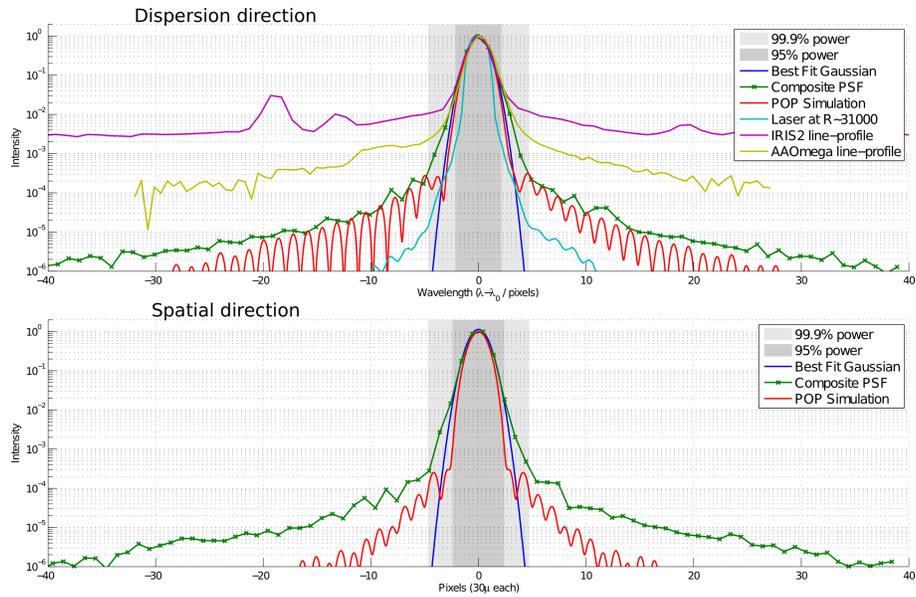

Fig. 3 – PSF line profiles (summed along the perpendicular axis with a Gaussian weighting, as would be done with real spectra) from PIMMS IR with comparisons in the spectral (top) and spatial (bottom) direction. Shown are: Green (stars) – composite of underexposed and overexposed profiles as explained in text; Red – ZEMAX POP simulation; Blue – Best fit Gaussian; Cyan – OSA spectrum of laser source; Purple and Yellow – Typical line profiles attributed to the IRIS2 150um slit and AAOmega fiber inputs respectively (The spectra from these spectrographs have been scaled to have the same FWHM determined from the best fit Gaussian, due to the difference in actual spectral resolution and pixel sampling). Note that the comparison spectra are shown to display the intensity level that wings of the respective PSF present.

*3.3 Photonic lantern input*

We then fed the PIMMS IR spectrograph with a nominally 7-mode PL device in order to measure the performance of a photonic lantern feed. To our knowledge, no other PL demonstrations have obtained spectra simultaneously of the complete PL feed. Studies of the spectral variability took spectra separately using an OSA [22]. A previous demonstration using an AWG-based PIMMS only used 12 of 19 of the SM ports of that PL, primarily due to physical space restrictions imposed by 125µm core spacing of conventional SMF and the need to maintain sufficient separation between cross-dispersed spectra [23, 24]. We overcame this by (i) using a smaller lantern, 1×7 vs. 1×19; (ii) forming the spectrograph entrance slit using the photonic TIGER configuration [11]. This allowed us to use conventional 125um fiber and form a compact slit for a small sacrifice in the effective spectral bandwidth.

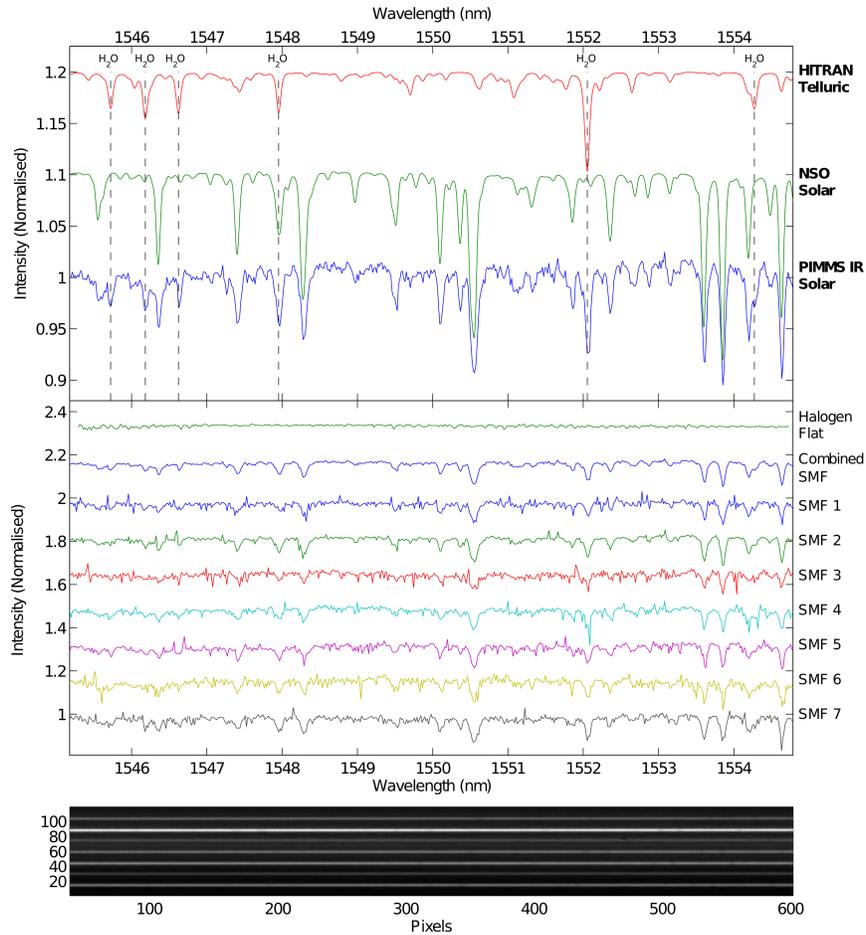

Fig. 4 – *Top:* Shown are: (i) A simulated atmospheric absorption spectrum using HITRAN [25]; (ii) Reference solar spectrum from NSO/Kitt Peak Observatory [26] convolved down to R~30000 for comparison. (iii) The PIMMS IR Solar spectrum obtained in Sydney, Australia. Spectra are offset vertically for convenience. The PIMMS IR spectrum shows the expected telluric absorption features (mostly water vapor) superimposed on the solar spectrum. *Middle:* From top to bottom the first spectrum is the response to a tungsten halogen lamp, the second is the solar spectrum obtained when combining the PL SMF outputs. The remaining seven are the individual spectra of the seven SMF outputs and their combination. Each spectrum is normalized by its respective median value, and offset for clarity. The noise level varies between spectra due to the differences in mean signal. *Bottom:* Detector image of the dispersed single-mode output fibers of the PL.

To obtain a final spectrum three sets of data (composed of light and dark frames) were acquired. The first was used for wavelength calibration, where the MM port of the PL is illuminated using the far-field of a SMF that is coupled to our tunable laser source. In the second, the MM port was directly illuminated with a tungsten halogen light source (Thorlabs OSL1-EC) to obtain a flat field response. Finally, raw spectra were obtained by aiming the MM port directly at the Sun. The individual spectra are shown in the middle part of Fig. 4, where each spectrum is normalized and offset. Also shown here is the flat response of PIMMS IR from a halogen source and the final combined spectrum formed from the average of the individual spectra. The noise visibly changes in the seven individual spectra; this is consistent with the relative intensity difference between each of them. In contrast to previous results [22], no spectral variation is seen at narrow bandwidth and high resolution between the individual spectra of the SMF outs of the 1×7 PL. Averaging the 7 individual spectra forms the complete lantern spectrum. This is shown in the top of Fig. 4 along with a reference solar spectrum [26] and a model telluric spectrum [25]. Both the comparison spectra were convolved with a Gaussian to have a similar resolution to the PIMMS IR spectrum. In the PIMMS IR wavelength range water vapor is the dominant absorber, the six strongest lines are marked in Fig. 4.

The throughput from slit to detector is simply measured using a power meter at both ends and illuminating the MM port with the tunable laser source. The measured throughput of the optics and diffraction grating is ~74%. Including a nominal detector QE of ~80% the spectrograph throughput is ~60%. The typical throughput of the current generation of 1×7 lanterns is ~85%. Thus the effective throughput, of PIMMS IR, including the lantern transition is ~50%. Performance could be improved with better-matched AR coatings on all optical surfaces.

## 4. Conclusion

Using a MM to SM converter known as a photonic lantern, we have demonstrated the power of the PIMMS concept. The photonic lantern allows us to remap an arbitrary input to an array of SMFs at the spectrograph's entrance slit, thus resulting in a diffraction-limited slit. This results in the most compact spectrograph possible (for a given resolving power) limited only by detector sampling requirements. More importantly, the spectrograph's optical design is completely decoupled from original light source. This means that the same design could in principle be used on a 30 cm telescope, a 30 m telescope or any other source that can be coupled to an MMF. The PIMMS IR detector requires the same area of an MM design, but for an MMF with $N$ (unpolarized) modes, we require $N$ SMFs. This demands at least $\varepsilon.N$ more pixels to appropriately sample the PIMMS spectra. For a conventional detector with square pixels, this leads to an $(\varepsilon.N)^{1/2}$ increase in read noise (assuming the smaller pixels have the same read noise as the larger pixels in the MM case) when the final spectra are combined. This increased noise in the system could be a limitation in extremely low-level light applications. However, this is partially balanced by the decreased size of the instrument, increases in throughput over conventional designs and the trend toward lower read noise detectors. In a subsequent paper, we show that the read noise penalty can be mitigated with a customized detector employing rectangular pixels. An initial design of such a device is under development with e2v (UK).

The PSF of the SMF pseudo-slit also shows an additional advantage of the photonic lantern approach. Scattered light is significantly reduced with <1% of the power outside the core of the PSF. This could be improved further using a customized optical design rather the COTS components. We also showed the photonic lantern faithfully reproduces the input source spectrum when combining the individual source spectra, and importantly does not appear to be affected by any wavelength dependent variation in intensity between SMF outputs.

## Acknowledgments

This research was supported by JBH through a Federation Fellowship from the Australian Research Council. Authors would also like to thank Jon Lawrence at AAO for use of equipment and Joel Rafael Salazar Gil for assistance in the fabrication of the 1×7 photonic lantern.